\documentstyle[prb,aps,epsf, preprint]{revtex}
\begin{document}
\title{Density of States in Superconductor -Normal 
Metal-Superconductor Junctions}
\author{F. Zhou$^{1,2}$,
P. Charlat$^{2}$,
B. Spivak$^{1}$,
B.Pannetier$^{2}$}
\address{$^{1}$Physics Department, University of Washington, Seattle, 
USA.\\ $^{2}$Centre de Recherches sur les Tr\`es Basses 
Temp\'eratures, C.N.R.S., Laboratoire associ\'e \`a l'Universit\'e 
Joseph Fourier, \\
F-38042 Grenoble, France}

\maketitle

\begin{abstract}
We consider the $\chi_0$-dependence of the density of states inside 
the normal metal of a superconductor - normal metal -
superconductor ($SNS$) junction.
Here $\chi_0$ is the phase difference of two superconductors of the 
junction. It is shown that in the absence of
electron-electron interaction the energy dependence of the density of 
states has a gap
which decreases as $\chi_0$ increases and closes at $\chi_0=\pi$. 
Both the analytical expressions for the $\chi_0$ dependence of the 
density of states and the results of numerical simulations are 
presented. \end{abstract}
PACS index category: 05.20-y, 82.20-w

\section{Introduction}¥
The proximity effect in a normal metal near superconductor-normal 
metal boundary has been studied for many years. Recently, progress in 
microfabrication technology has revived the interest in this 
phenomenon. Due to Andreev reflections at superconductor-normal metal boundaries,
the density of states inside the normal metal of a superconductor-normal metal 
($SN$) junction is different from that of the bulk normal metal. This phenomenon has 
been studied both theoretically and experimentally
\cite{Golubov95,Zirnbauer96,Frahm96,Melsen96,Gueron96}.
 In the superconductor - normal metal - superconductor ($SNS$) junction it has been 
shown that at $\chi_{0}=0$ the Andreev reflections at the 
superconductor-normal metal interfaces cause the density of states to 
have a gap, which is of the order of the Thouless energy $E_c ={D}/{L^2}\ll 
\Delta_0$. Here $\chi_0$ is the phase difference between the two 
superconductors,
$\Delta_0$ is the modulus of the order
parameter in the superconductors, $L$ is the length of the normal 
metal of the junction (see Fig.1) and $D$ is the diffusion constant 
in the metal.

In this paper we study the $\chi_0$ dependence of the density of 
states $\nu(\epsilon, \chi_0)$ in the case when the length of the 
normal metal region is much larger than the electron elastic mean 
free path (diffusive regime). We study both analytically and numerically 
the behaviour of the density of states near the gap edge. This quantity 
can be measured for example in a tunneling experiment. We will show 
below that it also determines the conductance of the junction $G_{SNS}$ at low 
temperatures $T$.

The density of states in the normal metal can be expressed in term of 
the retarded Green's function
$g^R(\epsilon, x)=\cos\theta(\epsilon,x)$\cite{Zhou95} 
\begin{equation}
\nu(\epsilon)=
\frac{\nu_0}{L}\int_{-L/2}^{L/2}\cos\theta_1(\epsilon, x) 
\cosh\theta_2(\epsilon, x)dx.
\end{equation}
Here $\nu_0=mp_F$ is the density of states in the bulk normal metal 
and $p_F$ is the Fermi wave length;
$\theta=\theta_1 +i\theta_2$ and $\chi=\chi_1 + i\chi_2$ are the 
complex variables described by Usadel equations \cite{Zhou95}, 
\begin{eqnarray}
\frac{D}{2}{\partial_x}^2\theta(\epsilon, x)+ 
i\epsilon\sin\theta(\epsilon,x)-\frac{D}{4}(\partial_x\chi(\epsilon,x))^2 
\sin2\theta(\epsilon, x)=\Delta_N(x)\cos\theta, \nonumber\\ 
\partial_x\{\partial_x\chi(\epsilon, x)\sin^2\theta(\epsilon, x)\}=0, 
\end{eqnarray}
\begin{equation}
\Delta_N(x)=\gamma_N \int d\epsilon \cos\theta_1(\epsilon, x) 
\sinh\theta_2(\epsilon,x)\tanh(\epsilon/2kT). \end{equation} 
$\Delta_{N}$ and $\gamma_N >0$ are the modulus of the order parameter 
and the dimensionless repulsive interaction constant inside the 
normal metal.

The boundary conditions for Eq.2 are (we consider the case when the 
transmission coefficient of the normal metal-superconductor boundary 
$t=1$),
\begin{eqnarray}
\theta(\epsilon, x=\pm\frac{L}{2})=\frac{\pi}{2}, \nonumber \\ 
\chi(\epsilon, x=\pm\frac{L}{2})=\pm\frac{\chi_0}{2}. \end{eqnarray}

Consider the situation when there is no electron-electron interaction 
in the normal metal and $\Delta_{N}=0$. We will see that in this case 
the energy gap $E_g(\chi_0)$ of the electron spectrum is a decreasing 
function of $\chi_0$ and is equal to zero only at $\chi_0=\pi$ (See 
Fig. 2). In two limiting cases, when
$\chi_0$ is close to $0$ or $\pi$ we have \begin{equation}
E_g(\chi_0)=E_c\left\{ \begin{array}{cc} C_2(1-C_1\chi^2_0) & 
\mbox{$\chi_0 \ll \pi$} \\ C_3(\pi -\chi_0) & \mbox{$\pi-\chi_0 \ll 
\pi$} \end{array} \right.
\end{equation}

\section{The energy gap}¥

To get this result we take into account that at $\epsilon< 
E_g(\chi_0)$, $\theta_1={\pi}/{2}$ following Eq.1. In this case the 
second integral of Eq.2 gives the solution for
$\theta_2(\epsilon, x)$:

\begin{eqnarray}
\int_{\theta_2(\epsilon, x)}^{\theta_{20}}d\theta_2 
\{\frac{\epsilon}{E_c}(\sinh \theta_{20}-\sinh\theta_2)+ 
\alpha^2(\epsilon)(\cosh^{-2}\theta_2-\cosh^{-2}\theta_{20})\}^{-\frac{1}{2} 
}
=\frac{2x}{L},
\end{eqnarray}
where $\theta_{20}(\epsilon)$ and $\alpha(\epsilon)$ are the 
functions of energy $\epsilon$ and $\chi_0$ determined by the 
equations \begin{eqnarray}
\int_{0}^{\theta_{20}}d\theta_2
\{\frac{\epsilon}{E_c}(\sinh \theta_{20}-\sinh\theta_2)+ 
\frac{\alpha^2(\epsilon)}{2}
(\cosh^{-2}\theta_2-\cosh^{-2}\theta_{20})\}^{-\frac{1}{2}} 
=1,\nonumber \\ \alpha(\epsilon)\int_{0}^{\theta_{20}}d\theta_2 
\cosh^{-2}\theta_2 \{\frac{\epsilon}{E_c}(\sinh 
\theta_{20}-\sinh\theta_2)+ \frac{\alpha^2(\epsilon)}{2}
(\cosh^{-2}\theta_2-\cosh^{-2}\theta_{20})\}^{-\frac{1}{2}} 
=\frac{\chi_0}{2}.\nonumber \\
\end{eqnarray}

One can show that Eq.7 has solutions only at low energy $\epsilon$. 
The energy gap $E_{g}(\chi_0)$ corresponds to the maximum value of 
$\epsilon$, at which a solution exists. Beyond this value, $\theta_{1} 
\neq {\pi}/{2}$ and therefore the density of states $\nu(\epsilon)$ is 
non zero.

Let us first consider the limiting case corresponding to $\chi_0 \ll \pi$. 
We can expand Eq.7 with respect to $\chi_0$ and as a result we have the 
following equation: 

\begin{eqnarray}
\sqrt{\frac{\epsilon}{E_c}}=
\int_{0}^{\theta_{20}}d\theta_2
(\sinh \theta_{20}-\sinh\theta_2)^{-\frac{1}{2}}- 
(\frac{\chi_0}{2})^2 A(\theta_{20}),
\end{eqnarray}
where $A$ is of the order of unity.
The right hand side of Eq.8 as a function of $\theta_{20}$ has a 
maxima equal to $\sqrt{C_2}(1-C_1\chi_0^2/2)$.
Here $C_{1,2}$ are numerical factors of order of unity. Their values 
can be obtained from the numerical solution of the 
Eq.2 \cite{Charlat97}: $C_{1}=0.91$ and $C_{2}=3.122$.
At low energies when $\epsilon \ll E_c$, Eq.8 has a solution 
$\theta_{20} \sim
{\epsilon}/{E_c}$ while at large energies $\epsilon >C_2 E_c(1 - C_1 
\chi^2_0)$,
Eq.8 does not have a solution. Therefore in this limit $E_{g}=C_2 E_c 
(1-C_1\chi^2_0)$, as given in Eq.5.

When $\chi_0-\pi\ll \pi$, the energy gap is small, $E_{g} \ll E_{c}$ 
and we can expand Eq.7 with respect to the small parameter ${\epsilon}/{E_{c}}$. 
As a result we have
\begin{equation}
\frac{\epsilon}{(\pi -\chi_0)E_c}=\pi^2\tanh\theta_{20}\{8B_1 
+(\pi-\chi_0)\pi B_2\sinh\theta_{20}\}^{-1}. \end{equation}
where $B_{1,2}$ are also constants of the order of unity. Again the 
right hand side of Eq.9 as a
function of $\theta_{20}$ reaches a maxima when \begin{equation}
\cosh\theta_{20} = B_{0}(\pi -\chi_0)^{-\frac{1}{3}} \end{equation}

Therefore, at $\epsilon > C_3E_c(\pi -\chi_0)$, Eq.9  has no
solution and $E_{g}(\chi_0)$ is given by Eq.5. From the numerical 
solution we find $B_{0}=2.21$ and $C_{3}=2.43$. The insert of Fig.2 
shows the linear dependence of the energy gap near $\chi_{0}=\pi$.

\section{The gap edge}¥

Let us turn to the calculations of the $\epsilon$-dependence of the 
density of states $\nu(\epsilon, \chi_0)$ at $\epsilon > E_{g}$.

In the region $\epsilon-E_g \ll E_g$, the quantities 

\begin{eqnarray}
\delta \theta_1(\epsilon, x)=\theta_1(\epsilon > E_g, x) 
-\frac{\pi}{2}, \nonumber \\
\delta
\theta_2(\epsilon, x)=\theta_2(\epsilon > E_g, x)-\theta_2(E_g, x), 
\nonumber \\
\delta
\chi_1(\epsilon, x)=\chi_1(\epsilon > E_g, x) -\chi_1(E_g, x), 
\nonumber \\ \delta \chi_2(\epsilon, x)=\chi_2(\epsilon > E_g, x), 
\end{eqnarray}
are small and go to zero as $\epsilon$ approaches $E_g$. Here 
$\theta_2(E_g, x)$, $\chi_1(E_g,x)$ are the solution of Eq.2 at 
$\epsilon =E_g$ given by Eqs.6,7. When $\pi -\chi_{0} \ll \pi$ 

\begin{eqnarray}
\sinh \theta_2(E_g,x)=\sinh \theta_{20}\cos\frac{\pi x}{L},\nonumber 
\\ \partial_x \chi_1(E_g, x)
=\frac{\pi \cosh\theta_{20}}{2L \cosh^2\theta_2(E_g,x)}, 
\end{eqnarray} where $\theta_{20}$ is given in Eq.10.

Expanding Eq.2 with respect to $\delta\theta_{1,2}$, 
$\delta\chi_{1,2}$ we get the following set of equations
\begin{eqnarray}
\{\frac{D}{2}\partial_x^2 + E_g \sinh\theta_2 + \frac{D}{2} (\partial_x \chi_1)^2\cosh 2\theta_2 \} \delta \theta_1 
-\frac{D}{2}\partial \chi_1 \sinh 2\theta_2 \partial_x \delta \chi_2  \nonumber \\
- \delta^3\theta_1 (\frac{E_g}{6}\sinh\theta_2 
+\frac{1}{3}D(\partial_x \chi_1)^2\cosh 2\theta_2) \nonumber \\
+\delta \theta_1 (E_g \cosh\theta_2 \delta \theta_2  
+ D(\partial_x\chi_1)^2\sinh 2\theta_2 \delta\theta_2+ D 
\partial_x\delta\chi_1 \partial_x\chi_1 \cosh 2\theta_2
-\frac{D}{2}(\partial_x \delta\chi_2)^2\cosh 2\theta_2) \nonumber \\
+\delta^2\theta_1 D \partial_x \delta\chi_2 \partial_x \chi_1 \sinh 
2\theta_2-D\partial_x\delta\chi_2 \delta \theta_2 \partial_x\chi_1 \cosh 
2\theta_2 -\frac{D}{2}\partial_x \delta \chi_2 \partial \delta \chi_1 
\sinh 2\theta_2  \nonumber \\
=-(\epsilon -\epsilon_g)\sinh\theta_2 \delta\theta_1 \\
\nonumber\\
\{\frac{D}{2}\partial_x^2 + E_g \sinh\theta_2 
+ \frac{D}{2} (\partial_x \chi_1)^2\cosh 2\theta_2 \} \delta \theta_2 
+\frac{D}{2}\partial_x \chi_1 \sinh 2\theta_2 \partial_x \delta \chi_1  \nonumber \\
- \delta^2\theta_1 (\frac{E_g}{2}\cosh\theta_2 
+ \frac{D}{2}(\partial_x \chi_1)^2\sinh 2\theta_2) 
+\delta\theta_1 D \partial_x \delta\chi_2 \partial_x \chi_1 \cosh 2\theta_2 \nonumber \\
-\frac{D}{4}(\partial_x \delta\chi_2)^2\sinh 2\theta_2 
=-(\epsilon -\epsilon_g)\cosh\theta_2 \\
\nonumber \\ 
\partial \{ \partial_x \delta \chi_2 \cosh^2\theta_2 
- \delta\theta_1 \partial_x 
\chi_1 \sinh 2\theta_2\}=0 \\
\nonumber \\
 \partial_x \{\partial _x \delta \chi_1 \cosh^2\theta_2 
 + \delta \theta_2 \partial_x \chi_1 \sinh 2\theta_2 
 - \delta \theta_1^2 \partial_x \chi_1\cosh 2\theta_2
  +\delta \theta_1 \partial_x \delta\chi_2 \sinh 2\theta_2 \}=0
\end{eqnarray}
with the boundary conditions for $\delta \theta_{1,2}, \delta 
\chi_{1,2}$: 

\begin{eqnarray}
\delta \theta_1(\epsilon, x=\pm\frac{L}{2})= \delta 
\theta_2(\epsilon, x=\pm\frac{L}{2})=0,
\nonumber\\
\delta \chi_1(\epsilon, x=\pm\frac{L}{2})= \delta \chi_2(\epsilon, 
x=\pm\frac{L}{2})=0. \end{eqnarray}

Consider the linear part of Eq.13-17 at $\epsilon-E_{g}=0$. Its solution 
determines the spatial dependence of $\delta\theta_{1,2}$, $\delta\chi_{1,2}$. 
Averaging Eq.13-17 over the sample and comparing the nonlinear 
terms in Eq.13 with right hand side term proportional to 
$(\epsilon-E_g)\sinh\theta_2 \delta\theta_1$, we can determine the 
value of $\delta\theta_1$ as a function of $\epsilon-E_g$. As a 
result we have 
\begin{equation}
\delta\theta_{10} \sim\sqrt{\frac{\epsilon -
E_g(\chi_0)}{E_c}{\cosh\theta_{20}}{C_4^2(\chi_0)}} 
\end{equation} 
where $C_4$ is a function of $E_g$ and consequently of $\chi_0$. 
Substituting Eqs.10, 18 into Eq.1 we obtain the asymptotic form valid 
when $\epsilon-E_{g}\ll E_g $:

\begin{equation}
\nu(\epsilon, \chi_0)=\nu_0 C_4(\chi_0)\sqrt{\frac{\epsilon- 
E_g(\chi_0)}{E_g(\chi_0)}}
\end{equation}
with $C_{4}\sim (\pi -\chi_0)^{-\beta}$, where $\beta\sim 2/3$.
 The value of $\beta$ is a result of a numerical calculation over 
 the whole energy range\cite{Charlat97}. The square root 
energy dependence of the density of states is illustrated in Fig.3
which shows the square of the density of states as function of the 
energy very close to the energy gap. At higher energy (Fig.3 insert), 
we find that the density of states exhibits a smooth bump above the gap when 
phase difference is small. This smooth maximum turns to
a sharper and sharper peak as the phase difference approaches $\pi$, 
i.e. as the gap closes.

It should be noted at this point that the asymptotic 
density of states at high energy or at $\chi_0=\pi$ is smaller than 
the normal state value. The region near the superconductor provides a 
small contribution to the (spatially averaged) density of states. This 
apparent deficit of states is balanced by the excess density of states 
above the superconducting gap $\Delta_0$ which according to our 
assumption is far above the energy range of interest. This deficit does 
not exist at the center of the normal metal or in a geometry where the 
relative area of the boundary between N and S goes to zero. 
The latter case is met in the billard geometry of Ref\cite{Melsen96}. 
 
\section{Discussion}¥

Although the above results were derived for the junction geometry shown in Fig.1, 
the $\chi_0$ and $\epsilon$ dependences of the density of states 
in Eqs.5,19 as well as the statement that the gap closes at 
$\chi_{0}=\pi$ are general\cite{levelspacing}. We believe that they are independent of
the transmission coefficient of the superconductor-normal metal 
boundaries and the geometry of the normal region. The values of $E_c$ 
and $C_i$, however,  depend on these parameters. Similar conclusions 
were also reached in \cite{Melsen96}. 

So far we assumed the electrons inside the normal metal do not 
interact with each other. In the presence of electron-electron 
interactions, $\Delta_N$ is not zero and one can study the 
interaction effects on the density of state
by taking into account $\Delta_N$ in Eq.7 for $E_g$. Since $\gamma_N 
\ll 1$, one can carry out the perturbative calculation with respect 
to $\gamma_N$. As a result, at $\chi_0 \ll \pi$, the gap turns out to 
be smaller than that given in Eq.5 for the noninteracting case, i.e. 
$E_g(0) -E_g(\gamma_N) \sim \gamma_N E_c$. Furthermore, the gap 
closes at $\chi^*$ smaller than $\pi$

\begin{equation}
\pi-\chi^* \sim \gamma_N
\end{equation}
Quantum fluctuations of the phase of the order parameter can also 
change the results derived above.

At finite temperature, in principle one also has to take into account 
the electron level broadening due to inelastic scattering. Such an 
inelastic process will introduce a temperature dependent density of 
state at the Fermi surface, i.e. $\nu(\epsilon=0)\sim 1/E_c 
\tau_{\epsilon}$. However at $T \ll E_{g}(\chi_0)$, the inelastic 
scattering rate $\tau_{\epsilon}^{-1}$ due to electron-phonon 
interaction is exponentially small.

Finally, let us calculate the conductance of the junction at  $T\ll 
E_{c}$. The conductance $G_{SNS}$ is the proportionality coefficient 
between the applied voltage and the disspative current averaged over 
the period of Josephson oscillations.
At low temperature, the main contribution to $G_{SNS}$ comes from the 
Debye relaxation mechanism\cite{Zhou97}. The qualitative picture is 
the following. When voltage $V$ is applied across the junction, the 
time dependence of $\chi_0$ is determined by the Josephson relation 

\begin{equation}
\frac{d\chi_0}{dt}=2eV
\end{equation}

At small $V$ the time dependence of $\nu$ is determined by the 
corresponding time dependence of $\chi_0(t)$. In other words the 
quasiparticle
energy levels move adiabatically with frequency $2eV$. The electron 
population of
the energy levels follows the motion of the levels and as a result 
the electron distribution becomes nonequilibrium. The relaxation of 
the nonequilibrium distribution due to inelastic processes leads to 
the entropy production and therefore contributes to the conductance. 
As a result, we have\cite{Zhou97},

\begin{equation}
G_{SNS}=\frac{e^2v}{\hbar^2 \nu_0}
\int d\epsilon \partial_\epsilon\tanh\frac{\epsilon}{2kT} 
\int_{0}^{2\pi}d\chi_0
{\tau_\epsilon(\chi_0)} \{\int^{\epsilon}_{-\infty} 
d\epsilon'\frac{d\nu(\epsilon', \chi_0)}{d\chi_0}\}^2 \end{equation} 
where $\tau_\epsilon(\chi_0)$ is the energy relaxation time of quasi 
particle of energy $\epsilon$. In this case\cite{Aronov81}, 

\begin{equation}
\frac{1}{\tau_\epsilon(\chi_0)}\sim\frac{\epsilon^3}{\Omega_D^2} 
\exp(-E_g(\chi_0)/k T)
\end{equation}
is exponentially small in the time interval when $E_g(\chi_0(t)) \gg 
T$. $\Omega_D$ is the Debye frequency. On the other hand, the 
concentration of quasiparticles in this case is also exponentially 
small. These two exponential factors cancel each other and the main 
contribution to Eq.22 comes from the time interval when 
$E_{g}(\chi_0(t))\sim T$. Since $\chi_0$ changes linearly with time 
and $E_{g}(\chi_0)$ vanishes linearly
as a function of $\chi_0-\pi$ at $\chi_0$ close to $\pi$, following 
Eq.22 we have

\begin{equation}
G_{SNS} \approx G_N C^2_4(\frac{T}{E_c}) \tau_{in}T
\end{equation}
Here $G_N=e^2 D \nu_0\frac{S}{L}$
and $\tau^{-1}_{in} \sim T^3/\Omega_D^2$. $C_4$ is given in Eq.19. At 
$E_{c}\sim T$, Eq.24 matches the result of the conductance, 
$G_N\tau_{in}E_c^2/T $, which was obtained in\cite{Zhou97} at the 
high temperature limit $T \gg E_c$.

Following the above arguments, the time dependence of the conductance 
has the form of narrow peaks with the amplitude of order of $G_N 
C^2_4(T/E_c)\tau_{in} E_c$ and duration of the order of 
$(eV)^{-1}{T}{E_{c}}$. This phenomenon can be connected with the well 
known $\cos\chi_0$ problem which has been investigated
both experimentally and theoretically \cite{Pedersen72,Zorin79}. 

\section{Conclusion}¥

In conclusion we have shown that the density of states in the normal 
part of a SNS junction has a gap which closes when the superconducting 
phase difference is $\chi_{0}=\pi$. The energy dependence of the 
density of states near the gap edge has been calculated.¥
We should mention that nonequilibrium effects in superconductors
which are connected with time dependence of quasiparticle spectrum 
have been considered in\cite{Aslamasov76,Schmid80}. In these papers 
the enhancement of the critical current due to the nonequilibrium effects was considered. 
The effect considered there is proportional to 
$(eV\tau_{\epsilon})^2$ while the $DC$ current calculated above is 
linear in $(eV\tau_{\epsilon})$.

We would like to acknowledge discussions with H. Courtois. 
This work was partially supported by the NATO CRG 960597.

\begin{figure}
\begin{center}
\leavevmode
\epsfbox{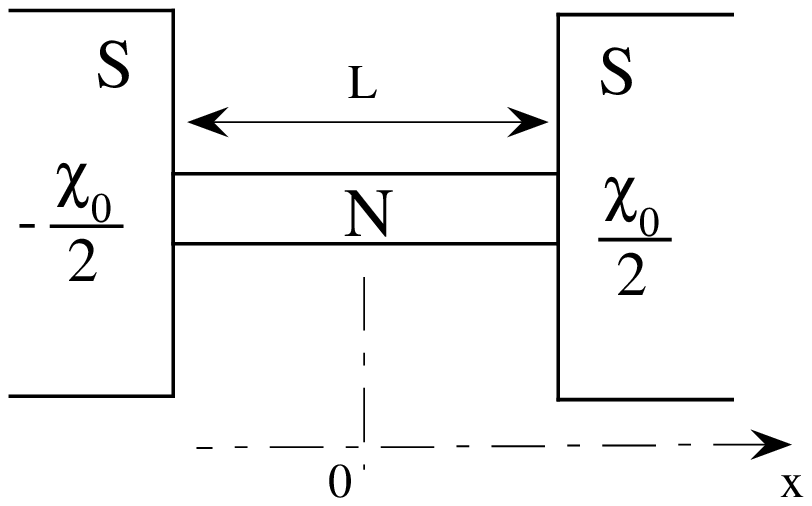}
\end{center}
\caption{
The S-N-S system: the origin of coordinate x is the center of 
the normal metal (length L). The phases of the superconducting order 
parameters in the right and left superconducting electrodes are 
respectively $-\chi_{0}/2$ and $+\chi_{0}/2$. 
}
\end{figure}

\begin{figure}
\begin{center}
\leavevmode
\epsfbox{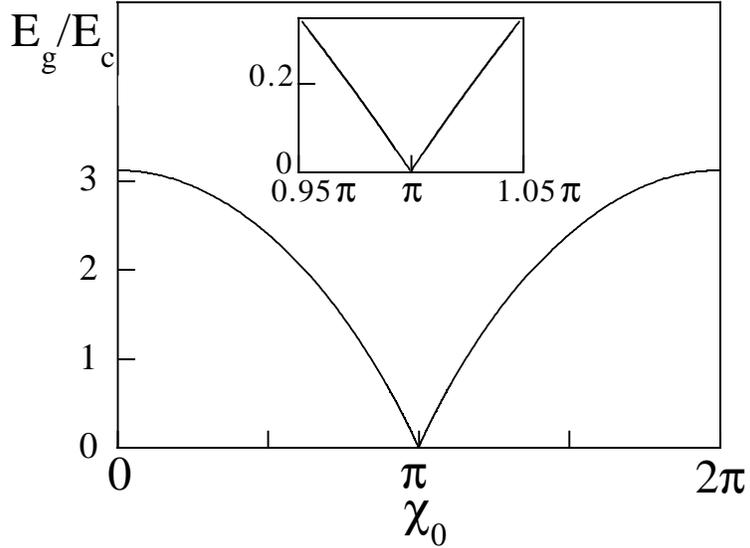}
\end{center}
\caption{
Energy gap $E_{g}$ in units of the Thouless energy $E_{c}$ vs 
the phase difference $\chi_{0}$ between superconducting contacts. The 
insert shows the linear dependence of $E_{g}$ near $\chi_{0}=\pi$. 
}
\end{figure}

\begin{figure}
\begin{center}
\leavevmode
\epsfbox{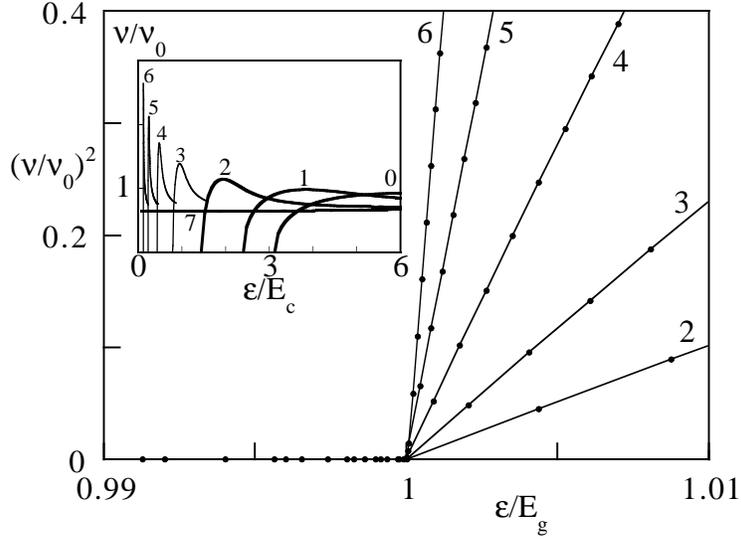}
\end{center}

\caption{
The square power of the density of states vs reduced energy 
$\epsilon/E_{g}(\chi_{0})$ near the gap edge. Curves labeled $0$, 
$1$, $2$, $3$, $4$, $5$, $6$ and $7$ are for $\chi_{0}=0$, $\pi/2$, 
$3\pi/4$, $7\pi/8$,
$15\pi/16$, $31\pi/32$,
$63\pi/64$ and $\pi$.
The insert shows the full density of states curves (here 
$E_{g}\ll\Delta_{0}$).
}
\end{figure}

\begin{figure}
\begin{center}
\leavevmode
\epsfbox{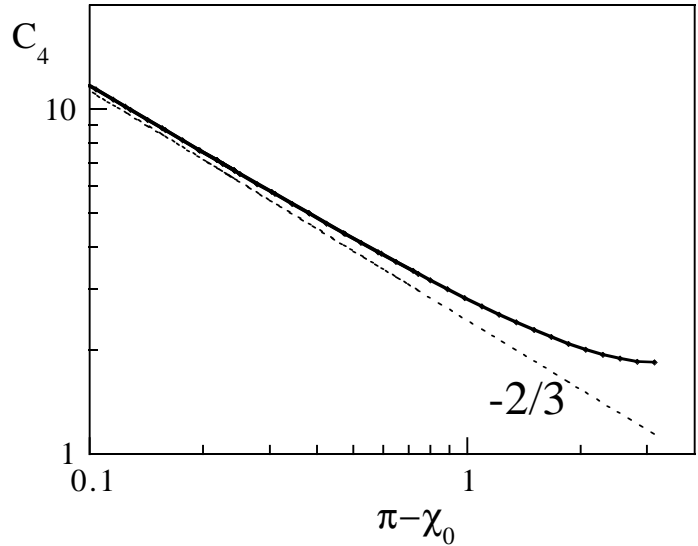}
\end{center}

\caption
{
Divergence of the $C_{4}$ numerical coefficient near 
$\chi_{0}=\pi$. The dashed line is an asymptotic law 
$(\pi-\chi_{0})^{-\beta}$ with $\beta=2/3$. 
}
\end{figure}

\end{document}